\documentclass[11pt]{cernrep}
\usepackage{axodraw,graphicx,cite}

\begin{document}

\def\x{\chi}
\def\ti{\tilde}
\def\nt{\tilde \x^0}
\newcommand{\mnt}[1]   {m_{\tilde\x^0_{#1}}}

\def\sigv{\langle\sigma v\rangle}
\def\sigva{\langle\sigma_A v\rangle}



\title{LHC\,/\,ILC\,/\,Cosmology Interplay 
   \footnote{ Contribution to the proceedings of the IX Workshop 
       on High Energy Physics Phenomenology (WHEPP-9),\\ 
       3--14 Jan 2006, Bhubaneswar, India.}}

\author{Sabine Kraml}
\institute{Theory Division, Dept.\ of Physics, 
         CERN, CH-1211 Geneva 23, Switzerland}

\maketitle


\begin{abstract}
There is a strong and growing interplay between particle 
physics and cosmology. In this talk, I discuss some aspects of this interplay 
concerning dark matter candidates put forth by theories beyond 
the Standard Model. In explaining the requirements for collider 
tests of such dark matter candidates, I focus in particular on the 
case of the lightest neutralino in the MSSM.
\end{abstract}

\section{Introduction}

Cosmological data ranging from rotation curves of spiral galaxies 
to the cosmic microwave background tell us that most of the mass 
in the Universe is provided by non-luminous, ``dark'' matter (DM); 
see \cite{Kolb:1990vq,Jungman:1995df,Bertone:2004pz} for reviews. 
To be concrete, the recent measurements from WMAP
\cite{Bennett:2003bz,Spergel:2003cb} and SDSS
\cite{Tegmark:2003ud} imply a (dominantly cold) dark matter
density of
$\Omega h^2 \simeq 0.1$ to an accuracy of about 10\% at $1\sigma$.  
The nature of this dark matter is one of the big puzzles of 
present-day physics. 
Although there are also other explanations, such as  
primordial black holes, many lines of reasoning suggest that the DM 
consists of a new weakly interacting massive particle, a so-called WIMP.

At the same time, we know that the Standard Model (SM) 
of particle physics, despite its tremendous success at energies 
up to $\sim 100$~GeV, is incomplete. For well-founded theoretical 
reasons, which I do not have to explain to this audience, we expect new 
physics beyond the Standard Model (BSM) to emerge at the TeV energy scale. 
This exciting new frontier will soon be probed by the LHC! 
In attempts to embed the SM in a more fundamental frame, theorists have 
come up with a wealth of BSM theories. These theories typically predict 
new particles, which may be discovered at the LHC. The lightest of these 
new particles is often stable by virtue of a new discrete symmetry  
and provides a natural DM candidate.

The dark matter candidates such put forth by particle physics are quite 
numerous \cite{Bertone:2004pz} and contain, for example, 
the lightest supersymmetric particle (LSP) 
in supersymmetry (SUSY) with R-parity conservation;  
the lightest Kaluza--Klein (KK) excitation in models with extra 
dimensions and KK-parity;  
the lightest T-odd state in little Higgs models with T-parity; etc. 
Note that all these possibilities 
are generally testable in collider experiments.
This creates a very strong interplay between particle physics, 
at both theoretical and experimental levels, and cosmology.

In this talk, I discuss this interplay using the example of the  
minimal supersymmetric standard model (MSSM) with a neutralino LSP  
as the DM candidate.
I focus in particular on the requirements for collider tests for  
determining whether the neutralino can be the DM in the Universe. 
I want to stress, however, that the general statements of this talk 
hold also for other, including non-SUSY, models.  

\section{Relic density of a WIMP}

\begin{figure*}[t]
\centering
\vspace*{2mm}
\includegraphics[height=5.5cm,width=5.5cm]{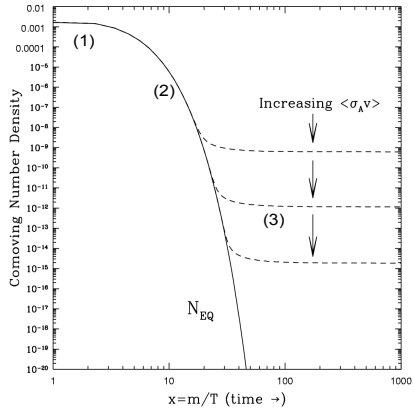}
\vspace*{2mm}
\caption{The cosmological evolution of a thermal relic's
  comoving number density, from~\protect\cite{Kolb:1990vq}. 
  The full line is the equilibrium abundance; the dashed lines 
  are the actual abundance after freeze-out. As the annihilation cross 
  section $\sigva$ is increased, the WIMP stays in equilibrium longer, 
  leading to a smaller relic density.}
\label{fig:freezeout}
\end{figure*}

The standard assumption is that the dark matter particle, 
let us call it $\chi$, is a thermal relic of the Big Bang as 
illustrated in Fig.~\ref{fig:freezeout}. 
The argument goes as follows. 
When the early Universe was dense and hot, $T\gg m_\chi$, 
$\chi$ was in thermal equilibrium;   
annihilation of $\chi$ and $\bar\chi$ into lighter particles, 
$\chi\bar\chi\to l\bar l$,
and the inverse process $l\bar l\to \chi\bar\chi$ 
proceeded with equal rates. 
As the Universe expanded and cooled to a temperature $T<m_\chi$, 
the number density of $\chi$ dropped exponentially, $n_\chi\sim e^{-m_\chi/T}$. 
Eventually the temperature became too low for the annihilation to 
keep up with the expansion rate and $\chi$ `froze out' with  
the cosmological abundance observed today. 

The time evolution of the number density $n_\chi (t)$ is described 
by the Boltzman equation,
\begin{equation}
  {\rm d}n_\chi/{\rm d}t + 3Hn_\chi = 
  - \sigva \, [(n_\chi)^2-(n_\chi^{\rm eq})^2]\,,
\end{equation}
where $H$ is the Hubble expansion rate, $n_\chi^{\rm eq}$ is the 
equilibrium number density, and $\sigva$ is the thermally averaged 
annihilation cross section summed over all contributing channels. 
It turns out that the relic abundance today is inversely 
proportional to the thermally averaged 
annihilation cross section, $\Omega_\chi h^2\sim 1/\sigva$. 
When the properties and interactions of the WIMP are known, its thermal 
relic abundance can hence be computed from particle physics' principles 
and compared with cosmological data.  

\section{Collider tests of BSM DM candidates}

As mentioned above, BSM scenarios with WIMP-DM candidates 
are testable at colliders. 
An important point in this regard is the following. 
Being electrically neutral and stable, the WIMP escapes
the detector as missing energy and momentum.
The preferred discovery channels therefore rely on the production
of other new particles present in the theory and their subsequent 
decays into the DM candidate. By measuring the properties and decay
kinematics of these new particles, one should then be able to
determine the properties of the WIMP. If the measurements are
precise enough, this allows to predict the annihilation cross
sections and hence the thermal relic density of the DM
candidate, thus checking the consistency between a particular
model of new physics and cosmology.

At the LHC, the generic WIMP signature is jets (plus leptons) plus 
large missing transverse energy. This is often regarded as the golden 
SUSY signature, but it also holds for other models. 
This is great for discovery, but resolving the underlying theory and 
the nature of the DM candidate will not be trivial. 
For mass measurements, precisions of a few percent can be expected. 
To decisively distinguish, for instance, SUSY from other BSM, spin 
measurements will be necessary. See \cite{Barr:2004ze} 
for first attempts along this line. 
In terms of predicting $\Omega_\chi h^2$, the precisions achievable at the 
LHC will, however, in general not be sufficient to match those of WMAP 
and other cosmological experiments. 
To this aim, precision measurements of masses, couplings and quantum 
numbers of the new states at a future International 
$e^+e^-$ Linear Collider (ILC) will be necessary in addition. 
This will be described in more detail below.

\section{Neutralino dark matter}

\subsection{Neutralino properties and annihilation channels}
\label{sect:neutprop}

Let me now put the spotlight on a particular BSM DM candidate, 
the neutralino LSP of the minimal supersymmetric standard model (MSSM).
The neutralino mass matrix in the bino--wino--higgsino basis 
$\psi_j^0=(-i\lambda ',\,-i\lambda^3,\,\psi_{H_1}^0,\,\psi_{H_2}^0)$ is
\begin{equation}
  {\cal M}_N =
  \left( \begin{array}{cccc}
  M_1 & 0 & -m_Z s_W c_\beta  & m_Z s_W s_\beta \\
  0 & M_2 &  m_Z c_W c_\beta  & -m_Z c_W s_\beta  \\
  -m_Z s_W c_\beta & m_Z c_W c_\beta   & 0 & -\mu \\
   m_Z s_W s_\beta & - m_Z c_W s_\beta & -\mu & 0
  \end{array}\right)
\label{eq:ntmassmat}
\end{equation}
with $M_{1}$ and $M_{2}$ the U(1) and SU(2) gaugino masses and 
$\mu$ the higgsino mass parameter. Furthermore,  
$s_W=\sin\theta_W$, $c_W=\cos\theta_W$, $s_\beta=\sin\beta$,
$c_\beta=\cos\beta$ and $\tan\beta = v_2/v_1$
($v_{1,2}$ being the vacuum expectation values of the two Higgs
fields $H_{1,2}$). 
This matrix is diagonalized by a unitary mixing matrix $N$,
\begin{equation}
  N^*{{\cal M}_N} N^\dagger =
  {\rm diag}(\mnt{1},\,\mnt{2},\,\mnt{3},\,\mnt{4})\,,
\end{equation}
where $\mnt{i}$, $i=1,...,4$, are the (non-negative) masses
of the physical neutralino states with $\mnt{1}<...<\mnt{4}$.
The lightest neutralino is then decomposed as
\begin{equation}
  \nt_1= N_{11}\tilde{B}+ N_{12} \tilde{W} +N_{13}\tilde{H_1}+
  N_{14}\tilde{H_2} \,.
\end{equation}
It will hence be mostly bino, wino, or higgsino, according to
the smallest mass parameter in Eq.~(\ref{eq:ntmassmat}),
$M_1$, $M_2$, or $\mu$, respectively. 
If the $\nt_1$ is the LSP and R-parity is conserved, this indeed provides 
an excellent cold dark matter candidate~\cite{Goldberg:1983nd,Ellis:1983ew}. 

\begin{figure}[htbp]
\centering 
{
\unitlength=1.0 pt
\SetScale{1.0}
\SetWidth{0.76}      
\normalsize    
{} \qquad\allowbreak

\begin{picture}(95,80)(0,0)
\Text(14.0,70.0)[r]{$\ti\chi_1^0$}
\Line(16.0,70.0)(58.0,70.0) 
\Text(82.0,70.0)[l]{$\bar{f}$}
\ArrowLine(79.0,70.0)(58.0,70.0) 
\Text(53.0,60.0)[r]{$\tilde{f}$}
\DashArrowLine(58.0,70.0)(58.0,50.0){1.0} 
\Text(14.0,50.0)[r]{$\ti\chi_1^0$}
\Line(16.0,50.0)(58.0,50.0) 
\Text(82.0,50.0)[l]{$f$}
\ArrowLine(58.0,50.0)(79.0,50.0) 
\end{picture} \quad
%
\begin{picture}(100,45)(0,0)
\Text(14.0,70.0)[r]{$\nt_1$}
\Line(16.0,70.0)(58.0,70.0) 
\Text(82.0,70.0)[l]{$W^-$}
\DashArrowLine(79.0,70.0)(58.0,70.0){3.0} 
\Text(53.0,60.0)[r]{$\ti\chi_j^{+}$}
\ArrowLine(58.0,70.0)(58.0,50.0) 
\Text(14.0,50.0)[r]{$\nt_1$}
\Line(16.0,50.0)(58.0,50.0) 
\Text(82.0,50.0)[l]{$W^+$}
\DashArrowLine(58.0,50.0)(79.0,50.0){3.0} 
\end{picture} \quad
%
\begin{picture}(95,45)(0,0)
\Text(14.0,70.0)[r]{$\nt_1$}
\Line(16.0,70.0)(58.0,70.0) 
\Text(82.0,70.0)[l]{$Z^0$}
\DashLine(79.0,70.0)(58.0,70.0){3.0} 
\Text(53.0,60.0)[r]{$\ti\chi_k^{0}$}
\ArrowLine(58.0,70.0)(58.0,50.0) 
\Text(14.0,50.0)[r]{$\nt_1$}
\Line(16.0,50.0)(58.0,50.0) 
\Text(82.0,50.0)[l]{$Z^0$}
\DashLine(58.0,50.0)(79.0,50.0){3.0} 
\end{picture}

\begin{picture}(95,45)(0,0)
\Text(14.0,70.0)[r]{$\ti\chi_1^0$}
\Line(16.0,70.0)(37.0,60.0) 
\Text(14.0,50.0)[r]{$\ti\chi_1^0$}
\Line(16.0,50.0)(37.0,60.0) 
\Text(47.0,64.0)[b]{$Z^0$}
\DashLine(37.0,60.0)(58.0,60.0){1.0}
\Text(82.0,70.0)[l]{$t$}
\ArrowLine(58.0,60.0)(79.0,70.0) 
\Text(82.0,50.0)[l]{$\bar{t}$}
\ArrowLine(79.0,50.0)(58.0,60.0) 
\end{picture} \quad 
%
\begin{picture}(100,45)(0,0)
\Text(14.0,70.0)[r]{$\ti\chi_1^0$}
\Line(16.0,70.0)(37.0,60.0) 
\Text(14.0,50.0)[r]{$\ti\chi_1^0$}
\Line(16.0,50.0)(37.0,60.0) 
\Text(47.0,64.0)[b]{$A^0$}
\DashLine(37.0,60.0)(58.0,60.0){1.0}
\Text(82.0,70.0)[l]{$b$}
\ArrowLine(58.0,60.0)(79.0,70.0) 
\Text(82.0,50.0)[l]{$\bar{b}$}
\ArrowLine(79.0,50.0)(58.0,60.0) 
\end{picture} \quad
%
\begin{picture}(95,45)(0,0)
\Text(14.0,70.0)[r]{$\nt_1$}
\Line(16.0,70.0)(37.0,60.0) 
\Text(14.0,50.0)[r]{$\tilde{\tau}_1$}
\DashArrowLine(16.0,50.0)(37.0,60.0){1.0} 
\Text(47.0,64.0)[b]{$\tau$}
\ArrowLine(37.0,60.0)(58.0,60.0) 
\Text(82.0,70.0)[l]{$\tau$}
\ArrowLine(58.0,60.0)(79.0,70.0) 
\Text(82.0,50.0)[l]{$\gamma$}
\DashLine(58.0,60.0)(79.0,50.0){3.0} 
\end{picture}
}\vspace*{-1cm}
\caption{Examples of processes contributing to neutralino (co)annihilation.}
\label{fig:feyn}
\end{figure}
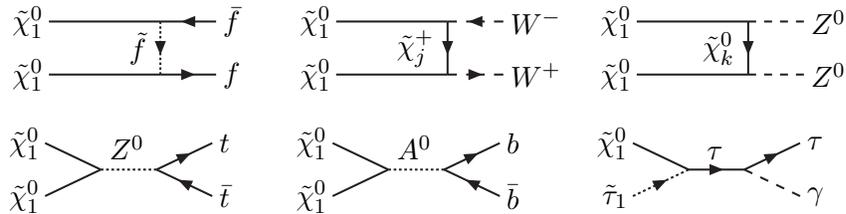

For its relic abundance to be in the right range, the neutralino LSP 
must annihilate efficiently enough. 
The relevant processes are (for a comprehensive discussion, 
see \cite{Drees:1992am}): 
annihilation of a bino LSP into fermion pairs through $t$-channel
sfermion exchange in case of very light sparticles;
annihilation of a mixed bino--higgsino or bino--wino LSP into gauge
boson pairs through $t$-channel chargino and neutralino exchange,
and into top-quark pairs through $s$-channel $Z$ exchange;
and finally annihilation near a Higgs resonance (the so-called Higgs funnel).
Furthermore, coannihilation processes with sparticles that are close
in mass with the LSP may bring $\Omega_{\ti\x} h^2$ in the desired range.
In particular, coannihilation with light sfermions can help to reduce
the relic density of a bino-like LSP.
The corresponding Feynman diagrams are depicted in Fig.~\ref{fig:feyn}.
Note that coannihilation generically occurs when there is
a small mass gap between the LSP and the next-to-lightest SUSY particle (NLSP).
In scenarios with a higgsino or wino LSP, one has in fact a mass-degenerate
triplet of Higgsinos or winos, and coannihilations are so efficient that
$\Omega_{\ti\x} h^2$ turns out much too small, unless the LSP has a mass of
order TeV.

Requiring that the neutralino LSP provides all of the cold dark matter, 
i.e. $0.095 < \Omega_{\ti\x} h^2 < 0.129$ at 2$\sigma$, puts strong 
constraints on the parameter space of the general MSSM. 
It puts even tighter bounds on models which assume specific relations 
between the soft-breaking parameters, such as mSUGRA, the CMSSM, AMSB, 
string-inspired models, etc.

\subsection{What do we need to measure? With which precision?}

\begin{figure}[t]
\centering
\vspace*{2mm}
\hspace*{6mm}
\includegraphics[height=5cm]{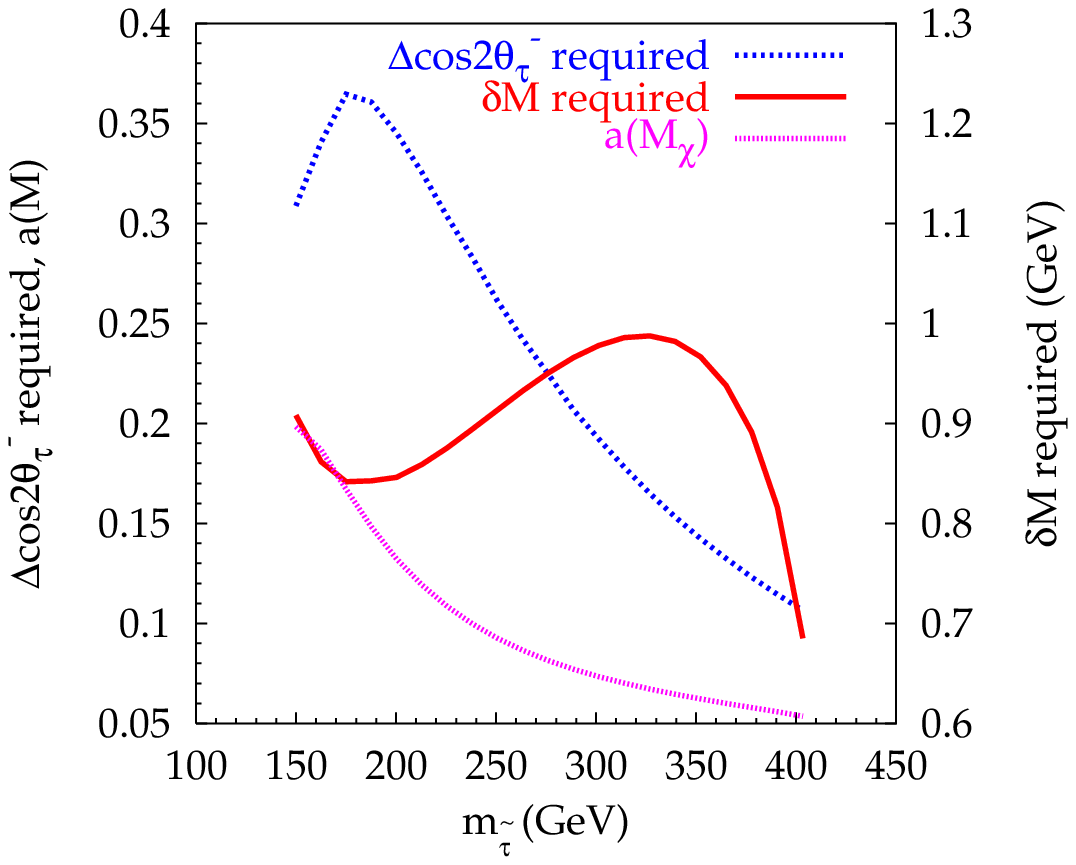}\hspace*{-4mm}
\includegraphics[height=5cm]{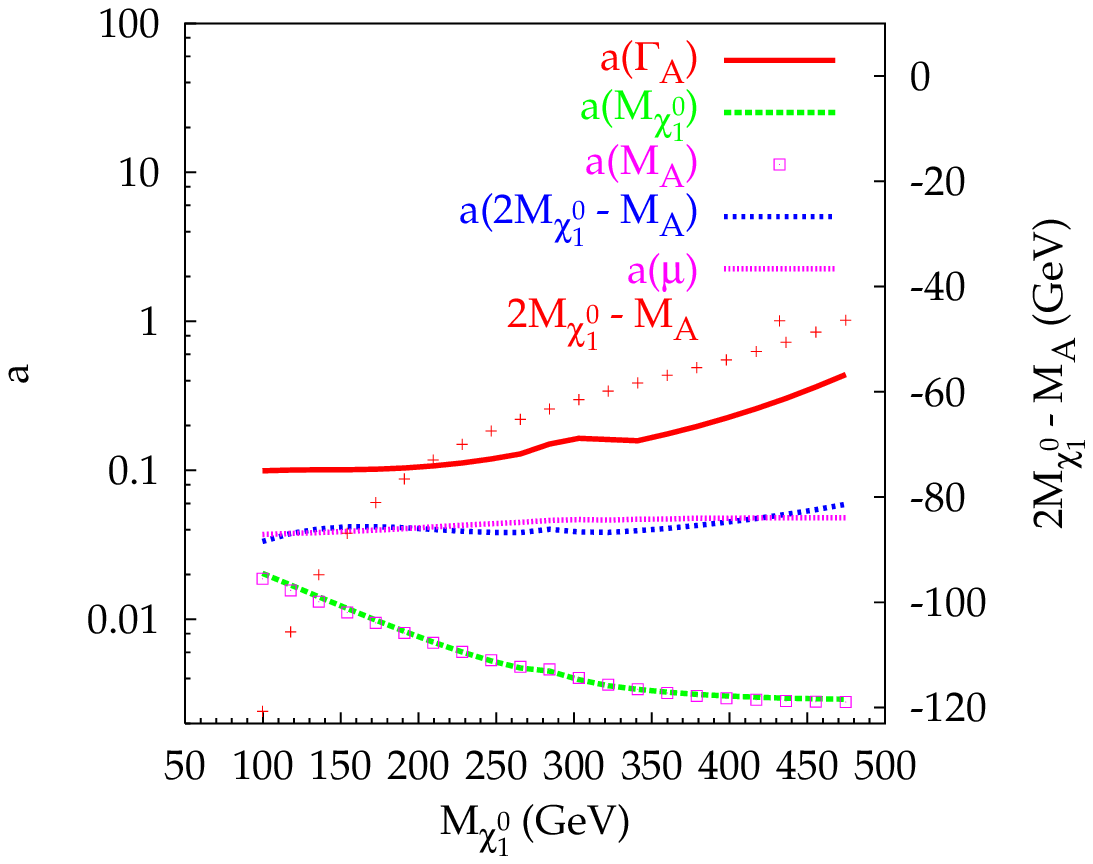}
\vspace*{2mm}
\caption{Fractional accuracies needed in the stau coannihilation (left) 
	and Higgs funnel (right) scenarios; from \protect\cite{Allanach:2004xn}.}
\label{fig:accuracies}
\end{figure}

From the above discussion it is clear that in order to predict 
the relic density of the LSP from collider experiments, 
we need to determine the properties of all sparticles 
potentially involved in the LSP (co)annhilation processes. 
That means not only determining the LSP mass but also the other neutralino, 
chargino, sfermion and Higgs masses (or at least lower limits on them 
to be sure that their contribution to $\sigv$ is negligible) 
as well as the relevant sparticle mixings. 
Naturally the question arises of which precisions are necessary in order 
to infer $\Omega_{\ti\x} h^2$ with $\sim 10\%$ accuracy, 
compatible with the WMAP precision. 
This was studied in detail in Ref.~\cite{Allanach:2004xn} 
for various cases of dominant (co)annihilation channels. 
Two examples, the case of stau coannihilation and the case of 
annihilation near the pseudoscalar Higgs resonance, are 
shown in Fig.~\ref{fig:accuracies}. 

Coannihilation with staus occurs for small neutralino--stau mass 
differences, typically 
$\Delta m_{\ti\x^0_1\ti\tau_1}=m_{\ti\tau_1}-m_{\ti\x^0_1}\le 10$~GeV. 
As can be seen in Fig.~\ref{fig:accuracies}, this mass difference needs 
to be measured to better than 1~GeV. 
Obviously this will be very difficult at the LHC, where one relies on 
the measurement of the $\tau^+\tau^-$ invariant mass from cascade decays 
involving 
\begin{equation}
  \ti\x^0_2 \to \tau^\pm\ti\tau^\mp_1 \to \tau^+\tau^-\ti\x^0_1\,.
\label{eq:tautau}
\end{equation}
The reason is that for small $\Delta m_{\ti\x^0_1\ti\tau_1}$ the 
$\tau$ from the $\ti\tau_1 \to \tau\ti\x^0_1$ decay is soft and can 
easily be missed. Moreover, even if both taus in (\ref{eq:tautau}) 
can be detected, one cannot tell whether the soft one comes from 
the $\ti\x^0_2 \to \tau\ti\tau_1$ or the $\ti\tau\to \tau\ti\x^0_1$ 
decay. 
Therefore one cannot tell whether it is $\Delta m_{\ti\x^0_1\ti\tau_1}$
or $\Delta m_{\ti\x^0_2\ti\tau_1}$ that is small. 
In addition to a precise determination of $\Delta m_{\ti\x^0_1\ti\tau_1}$, 
one also needs to measure the absolute masses and the mixing angles 
with good accuracy; see Fig.~\ref{fig:accuracies}. 
As shown in \cite{Bambade:2004tq}, this may be achieved in the clean 
experimental environment of the ILC; tunable beam energy and 
beam polarization are, however, essential for this aim. 

The Higgs funnel and the higgsino-LSP scenarios are even more challenging. 
In the former, one needs precisions at the percent or even permille level 
on $m_{\ti\x^0_1}$, $m_A$, the distance from the pole $m_A-2m_{\ti\x^0_1}$, 
the pseudoscalar width $\Gamma_A$, and the higgsino mass parameter $\mu$.
In the latter, one needs to resolve the complete neutralino system 
with $\sim 1\%$ accuracy. See \cite{Allanach:2004xn} for more details.
The limitation of the ILC lies mostly in the kinematic reach: the heavy 
Higgs and neutralino/chargino states may simply be too heavy to be 
produced, in which case one has to rely on LHC measurements. 
See \cite{Desch:2003vw} for an example of SUSY parameter determination 
from combined LHC/ILC analysis.

It becomes clear that for a precise prediction of $\Omega_{\ti\x} h^2$ 
from colliders, we will need precision measurements of most of the MSSM 
spectrum. A combination of LHC and ILC analyses \cite{Weiglein:2004hn} 
may prove essential for this purpose.

\subsection{What if CP is violated?}

So far we have only considered the CP-conserving MSSM. 
In general, however, several parameters of the MSSM can be complex, 
thus introducing new sources of CP violation (CPV) in the model.  
This is indeed a very interesting possibility since,    
given the Higgs mass bound of $m_h > 114$~GeV from LEP \cite{Barate:2003sz},
the amount of CPV present in the SM  
is not sufficient to generate the correct baryon asymmetry in 
the Universe. 

The parameters that can have CP phases are the gaugino
and higgsino masses and the trilinear sfermion--Higgs couplings.
Although constrained by electric dipole moments, non-zero phases can
significantly influence the phenomenology of SUSY particles.
They can also have a strong impact on the Higgs sector, inducing
scalar--pseudoscalar mixing through loop effects~\cite{Pilaftsis:1998dd}.
This can in turn have a potentially dramatic effect on the 
relic-density prediction in the Higgs funnel region:
neutralino annnihilation through $s$-channel scalar exchange is $p$-wave
suppressed; at small velocities it is dominated by pseudoscalar exchange.
In the presence of phases, both heavy Higgs bosons can, for instance,
acquire a pseudoscalar component and hence significantly contribute
to neutralino annihilation, even at small velocities. Likewise, when
only one of the resonances is accessible to the neutralino annihilation,
large effects can be expected by changing the scalar/pseudoscalar content
of this resonance. 

\begin{figure}[t]
\centering
\includegraphics[height=5.5cm]{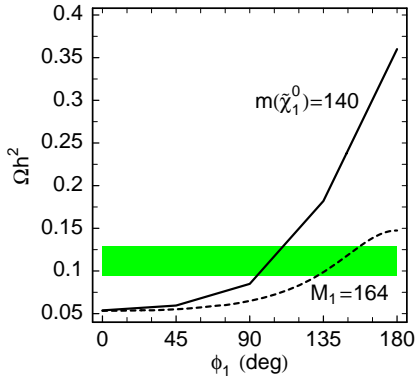}
\caption{$\Omega_{\ti\x} h^2$ as a function of $\phi_1$ for a mixed 
  bino--higgsino LSP scenario with $\mu=200$~GeV and $\tan\beta=10$; 
  the dashed line is for fixed $M_1=164$~GeV,
  while for the full line $M_1$ is adjusted such that $\mnt{1}=140$~GeV 
  remains constant.
  The green (grey) band shows the $2\sigma$ WMAP range.
  From \protect\cite{Belanger:2006qa}.}
\label{fig:mlsp140}
\end{figure}

CP phases however do not only impact $\Omega_{\ti\x} h^2$ in 
the Higgs funnel region --- they can lead to important effects in 
almost any scenario of neutralino (co)annihilation. 
In Ref.~\cite{Belanger:2006qa} we analysed in detail the influence 
of CP phases in the various scenarios of neutralino annihilation and
coannihilation for which the LSP is a `good' DM candidate. 
We found effects of up to an order of magnitude
from modifications in the couplings due to non-zero CP phases 
(note that care has to be taken to disentangle phase effects in 
couplings and kinematics).
As an example, Fig.~\ref{fig:mlsp140} shows $\Omega_{\ti\x} h^2$ 
as a function of the phase of $M_1$, $M_1=|M_1|e^{i\phi_1}$, 
for a mixed bino--higgsino LSP, which dominantly annihilates 
into $W^+W^-$. 

The largest effects are found, as expected, for annihilation through 
Higgs exchange, not only because of scalar--pseudoscalar Higgs mixing but 
also because a phase in the neutralino sector modifies the LSP's couplings 
to scalar and pseudoscalar states \cite{Belanger:2006qa}. 
Let me emphasize, however, that even in scenarios which feature a modest 
phase dependence, the variations in $\Omega_{\ti\x} h^2$ often exceed $10\%$. 
Therefore, when aiming at a precise prediction of the neutralino relic
density from collider measurements, one does not only
need precise sparticle spectroscopy as assumed in \cite{Baltz:2006fm} 
--- one also has to precisely determine the relevant couplings.
This includes the determination of possible CP phases.

\subsection{What if the inferred $\Omega h^2$ is too high?}

Suppose that we have discovered SUSY with a neutralino LSP and made 
precision measurements of all the relevant parameters, as pointed 
out above. It may turn out that the $\Omega_{\ti\x} h^2$ of the 
neutralino thus inferred is below the cosmological dark matter abundance. 
We would be lead to conclude that there exists yet another DM 
constituent, or else that there is some non-thermal mechanism 
contributing to the neutralino relic density. 

However, what if the inferred $\Omega_{\ti\x} h^2$ is too high?
There are several solutions to this problem. For one, the $\nt_1$ 
may only appear to be the LSP but decay 
into an even lighter sparticle outside the detectors. In this case the 
$\nt_1$ would actually be the NLSP and the real LSP would be, for instance, 
a gravitino or axino. The thermal relic abundance 
would then roughly scale with $m_{\rm LSP}/m_{\rm NLSP}$. 
Strong constraints on this scenario come from BBN limits on late 
$\nt_1\to Z\ti G$, $h\ti G$, etc., decays. 
(For details, see \cite{Feng:2005nz}.)

Solution number two is that R-parity is violated after all, but 
on long time scales so that the $\nt_1$ again appears stable in 
collider experiments. On cosmological scales, late decays of the 
$\nt_1$ would have reduced its number density to zero, and the actual 
dark matter would be something else. 

Solution number three is that our cosmological assumptions are wrong. 
Our picture of dark matter as a thermal relic from the Big Bang may well 
be too simple; the Early Universe may have evolved differently. 
Note also that the exact value and uncertainty of $\Omega h^2$ 
extracted from cosmological data depend on 
both the precise datasets used and the choice
of parameters allowed to vary \cite{Lahav:2006qy}.

\section{Conclusions}

We expect new physics beyond the Standard Model at the TeV energy 
scale, and there is well-motivated hope that this new physics also 
provides the dark matter of the Universe. This creates a strong 
interplay between theoretical particle physics, collider experiments, 
and cosmology. Using the example of a neutralino LSP in the MSSM, 
I have argued that precision measurements at LHC and ILC will be 
necessary to pin down the nature and properties of the dark matter 
candidate. 

Two more comments are in order. 
First, even if BSM with a WIMP DM candidate is discovered at 
future colliders, the properties of the WIMP are measured precisely 
and its $\Omega h^2$ thus deduced matches the cosmological value, 
this is no proof that this WIMP is indeed the DM. Direct and/or 
indirect detection \cite{Jungman:1995df,Bertone:2004pz} 
will be indispensible in addition. Here note 
that the rates for (in)direct detection can also be predicted once 
the properties of the WIMP are known. 
Second, the interplay between particle physics and cosmology is not 
limited to the dark matter question. Another example for this interplay 
is the baryon asymmetry of the Universe, where particle physics offers 
explanations through electroweak baryogenesis or leptogenesis. 

To conclude, I think it is fair to say that these are exciting times 
for both particle physics and cosmology, with a steadily growing 
synergy between the two fields.

\section*{Acknowledgements}

WHEPP-9 was a wonderful workshop, leaving a rich memory of physics 
discussions, old and new acquaintances with colleagues from India 
and elsewhere, strolls into the streets of Bhubaneswar, and of 
course Indian cuisine. I wish to thank the organizers for inviting 
me there. 
Financial support by the Austrian Academy of Sciences and 
the WHEPP-9 host organization, Institute of Physics, 
is also gratefully acknowledged.


\end{document}